# Les Générateurs de Scénarios Économiques : quelle utilisation en assurance ? [1]

**Alaeddine FALEH**[2]   **Frédéric PLANCHET**[3]   **Didier RULLIERE**[4]

**ISFA- Université Lyon I**[5]
**Caisse des Dépôts et Consignations**[6]

## RÉSUMÉ

Dans cet article, nous mettons en évidence les principales composantes d'un générateur de scénarios économiques (GSE) que ce soit au niveau de sa conception théorique ou au niveau de sa mise en œuvre pratique. Le choix de ces composantes est supposé être lié à la vocation finale du générateur de scénarios économiques que ce soit en tant qu'outil d'évaluation des produits financiers (*pricing*) ou en tant qu'outil de projection et de gestion des risques. Par ailleurs, nous développons une étude sur certains indicateurs de mesure de la performance du GSE comme un outil en amont du processus de prise de décision : à savoir la stabilité et l'absence de biais. Une application numérique permettant d'illustrer ces différents points est présentée à la fin.

MOTS-CLEFS : générateur de scénarios économiques, arbre de scénarios, stabilité, absence de biais

---

# ABSTRACT

In this paper, we present the principal components of an economic scenario generator (ESG), both for the theoretical design and for practical implementation. The choice of these components should be linked to the ultimate vocation of the economic scenario generator, which can be either a tool for pricing financial products or a tool for projection and risk management. We then develop a study on some performance measure indicators of the ESG as an input for the decision-making process, namely the indicators of stability and bias absence. Finally, a numerical application illustrates the main ideas of the paper.

# 1. Introduction

La projection sur le long terme des valeurs de marché des actifs financiers et des variables macro-économiques, souvent appelée « génération de scénarios économiques », constitue une phase cruciale dans la gestion actif-passif d'une compagnie d'assurance ou d'un fonds de retraite. Elle est un élément central de l'évaluation des provisions pour les garanties financières sur des contrats d'épargne dans le cadre de Solvabilité 2 (cf. Planchet [2009]). Dans le cas particulier de l'allocation stratégique[7] d'actifs, la détermination de l'allocation optimale vient dans un deuxième temps, ultérieurement à la projection des scénarios, afin de refléter l'attitude face au risque de l'investisseur de long terme (cf. Meucci [2005]).

Un générateur de scénarios économiques (GSE) s'avère ainsi un outil important d'aide à la décision dans le domaine de la gestion des risques en permettant d'obtenir des projections dans le futur des valeurs des éléments présents dans les deux compartiments du bilan de la société : les actifs (actions, obligations, immobilier,..) et le passif (provisions techniques, dettes financières,..). L'obtention de ces valeurs passe par la projection d'autres variables macro-économiques et financières telles que les taux d'intérêt, l'inflation des prix, l'inflation des salaires et le taux de chômage. A titre d'exemple, les prix futurs des obligations sont souvent déduits à partir de la projection des taux d'intérêt (cf. Ahlgrim et al. [2008]), de même les prestations futures de la société peuvent être indexées sur l'inflation des prix et/ou l'inflation des salaires (cf. Kouwenberg [2001]).

De manière générale, les GSE permettent de prendre en compte l'horizon long d'investissement des sociétés d'assurance et des fonds de retraite ainsi que l'influence des évolutions futures des variables macro-économiques et financières sur le choix de leurs paramètres techniques, tel que le taux de rendement garanti. Ceci est souvent motivé par la volonté de garantir un surplus permanent (l'écart entre la valeur des actifs financiers à une date donnée et la valeur des engagements actualisés à un taux d'intérêt de référence).

Du point de vue opérationnel, la mise en œuvre de ces projections s'est basée initialement sur des méthodes déterministes, dans des logiques de scénarios, pour tester le comportement des objectifs techniques dans différentes situations considérées comme caractéristiques. Grâce au développement concomitant de l'outil informatique et des techniques de simulation, il est devenu possible de générer à moindre coût de très nombreux scénarios en tenant compte des interactions entre les multiples sources de risque et de leurs distributions, dans une perspective probabiliste.

La construction et la mise en œuvre d'un GSE passent par les quatre étapes suivantes (voir par exemple Hibbert et al. [2001] ou Ahlgrim et al. [2005]) : la première étape consiste en l'identification des sources de risque prises en compte et des variables financières à modéliser (taux d'intérêt, inflation, rendement des actions, etc.) qu'on appellera dans la suite les variables du GSE. Ensuite est effectué le choix du modèle pour la dynamique de chacune de ces variables. La troisième étape consiste à sélectionner une structure de dépendance entre les sources de risque de façon à obtenir des projections cohérentes. Enfin l'estimation et le calibrage des paramètres des modèles retenus doit être effectuée. L'analyse des résultats obtenus de chaque GSE se fait ensuite en termes probabilistes en analysant la distribution d'indicateurs clé tels que le surplus, ou la valeur nette de l'actif.

---

[7] La caractérisation de « stratégique » vient simultanément de l'horizon temporel de long terme auquel s'appliquent cette allocation et du nombre limité de classes d'actifs considérées, généralement limité à 5 (ou au maximum à une dizaine).

Construire un GSE pertinent pour l'ensemble des problématiques techniques n'est pas chose facile, et en pratique les objectifs des décisionnaires en termes de choix de gestion ont une incidence sur la manière de structurer le générateur (cf. Ahlgrim et al. [2008]). On peut distinguer deux niveaux de difficultés : celui relatif à la conception théorique d'un GSE et celui relatif à sa mise en œuvre pratique. Il est donc particulièrement important de définir avec précision les principaux éléments qui caractérisent un GSE ainsi que les indicateurs de mesure de sa qualité et de sa performance. C'est l'objet de cette synthèse.

Après avoir recensé les principales structures théoriques des GSE proposés dans la littérature ainsi que les problématiques de conception correspondantes, on s'intéresse aux aspects relatifs à leur mise en œuvre. On présente ainsi les caractéristiques des différentes structures schématiques de projection de scénarios utilisées en pratique. Toujours dans le cadre de la mise en œuvre, on s'inspire des travaux de Mitra [2006] et de Zenios [2005] pour exposer les différentes méthodologies de génération des scénarios, en particulier les modèles ayant trait à la dynamique des variables du GSE. Ensuite, on présente une série d'indicateurs, qualitatifs et quantitatifs, pour la mesure de la qualité d'un GSE tout en étudiant leurs limites. Enfin, les principaux éléments exposés tout au long de l'article seront illustrés à travers une application numérique.

## 2. Présentation théorique des GSE

Dans ce qui suit, une première partie est consacrée aux problématiques souvent rencontrées en matière de conception théorique d'un modèle de GSE. La deuxième partie s'intéresse, quant à elle, à la revue des principaux travaux menés dans la littérature sur les GSE en les classifiant en fonction des choix effectués au niveau de la structure de dépendance entre les variables.

### 2.1. Introduction

Dans cette section, les problématiques théoriques liées à la construction du modèle sont exposées. Ces dernières traitent de questions qui permettent l'amélioration de la qualité des résultats et le développement du modèle pour l'adapter au contexte d'étude. Les problèmes de modélisation sont en premier lieu liés au choix de la représentation de la structure par terme des taux d'intérêt (STTI). La modélisation de la dynamique des rendements des actions ainsi que les autres variables suscite également des questions délicates.

En se basant sur les travaux de Roncalli [1998] et Décamps [1993], deux grandes classes de modèles de taux pour l'évaluation d'actifs financiers peuvent être présentées : les modèles d'absence d'opportunité d'arbitrage (AOA) et les modèles d'équilibre général.

L'évaluation dans le cadre des modèles d'AOA est de nature purement financière. Elle repose entièrement sur l'hypothèse d'absence d'opportunité d'arbitrage. L'utilisation de cette hypothèse fondamentale a permis de mettre en évidence deux approches d'évaluation par arbitrage :

- La première approche considère le prix des instruments financiers comme fonction de variables d'état. Ces variables d'état –généralement le taux court pour les modèles univariés (cf. Vasicek [1997]), ou le couple (taux court, taux long) pour les modèles bivariés (cf. Brennan et Schwartz [1982]) sont supposées de dynamique exogène. Dans de tels modèles, l'hypothèse d'absence d'opportunité d'arbitrage exprime que la prime de risque du marché[8] est indépendante de la maturité du titre considéré. Le prix d'un produit obligataire est ensuite obtenu comme solution d'une équation aux dérivées partielles. La résolution probabiliste de ces équations permet une meilleure interprétation financière des formules d'évaluation.

- La deuxième approche d'évaluation par arbitrage est celle proposée par Ho et Lee [1986] et Heath, Jarrow et Merton [1992]. Le point clef de cette approche est la prise en compte de toute l'information contenue dans la structure de taux initiale en considérant comme donnée exogène la dynamique simultanée de taux ayant différentes maturités appelée aussi dynamique des taux terme contre terme. Cette dynamique est choisie de façon à ce que l'hypothèse d'absence d'opportunité d'arbitrage soit respectée. Contrairement à l'arbitrage traditionnel la prime de risque du marché n'est pas spécifiée de façon exogène mais elle est définie implicitement par la dynamique des taux terme contre terme. Par ailleurs, aucune hypothèse sur la forme spécifique des prix des produits obligataires comme fonction d'une variable d'état n'est faite. Enfin, la dynamique de prix des zéro-coupons caractérise entièrement le modèle de courbe de taux.

L'évaluation dans le cadre d'un modèle d'équilibre général ne nécessite pas d'hypothèse sur les dynamiques de prix ou de taux. Ces dynamiques sont obtenues de manière endogène. Cette approche est celle de Cox, Ingersoll et Ross [1985] et de Campbel [1986]. En effet, les modèles d'équilibre typiques se basent sur les anticipations des mouvements futurs des taux d'intérêt de court terme et non pas sur la courbe de taux observé à la date initiale. Ces mouvements peuvent être donc dérivés à partir d'hypothèses plus générales sur des variables d'état qui décrivent l'ensemble de l'économie. En utilisant un processus des taux courts, on peut déduire le rendement d'une obligation de long terme en déterminant la trajectoire espérée des taux courts jusqu'à la maturité de cette obligation. L'intérêt d'une telle approche est triple :

- s'assurer que les processus étudiés sont cohérents avec un équilibre général,

- analyser la déformation de la courbe des taux en fonction des chocs sur les variables économiques sous jacentes,

- fournir une spécification fondée pour l'expression des prix ou des risques de marchés utilisés dans la valorisation par arbitrage.

Mais, les modèles d'équilibre de structure par terme génèrent des prix de produits de taux qui sont potentiellement incohérents avec ceux observés sur le marché à la date initiale. Même si les paramètres de ces modèles peuvent être calibrés avec une précision significative, la structure par terme résultante peut générer des prix éloignés de ceux observés sur le marché.

---

[8] Supplément de rendement exigé par les investisseurs pour avoir assumer le risque de détenir des actifs risqués plutôt que des actifs sans risque.

Le tableau suivant illustre les points clefs de ces modèles:

|  | Modèles d'Absence d'Opportunité d'Arbitrage (AOA) | | Modèles d'Equilibre Général |
| --- | --- | --- | --- |
|  | *Approches à prime de risque exogène* | *Approches à prime de risque endogène* | *Approches entièrement endogènes* |
| **Courbe des taux initiale** | - Utilisée. | - Utilisée. | - Non utilisée. |
| **Dynamique des taux** | Exogène pour des taux de maturité spécifié (court, long,..). | Exogène dans le cas des taux terme contre terme. | Endogène : Anticipation des taux court terme futurs à partir d'hypothèses générales sur les variables qui décrivent l'ensemble de l'économie. |
| **Prime de risque** | Exogène et indépendante de la maturité. | Endogène : déduite à partir de la dynamique des taux terme contre terme. | Endogène. |
| **Prix des produits de taux (obligations, etc.)** | - Prix fonction de variables d'état (taux courts, taux longs, etc.).<br><br>- Prix solution d'équations aux dérivés partielles. | - Aucune hypothèse sur la forme spécifique des prix comme fonction de variables d'état. | - Prix déterminés à partir des projections de taux courts. |

Une autre problématique théorique relative aux différentes variables modélisées, en particulier les taux d'intérêt, concerne le choix du nombre et de la nature des facteurs à utiliser pour la modélisation. Dans le cas particulier des taux d'intérêt, le choix définitif de ces deux éléments dépend du contexte d'application du GSE (projection de grandeurs réelles, évaluation de produits dérivés,…). La nature des facteurs peut être de sources différentes : elle peut être par exemple liée à l'horizon du taux (taux court, taux long) ou à la structure de la courbe des taux (facteur de courbure, de translation,…). A ce titre, Date et al. [2007] montre la significativité statistique du choix de deux facteurs (taux court, taux long) pour expliquer l'évolution de la courbe de taux.

Concernant les actions, différentes problématiques liées à l'appréhension de la dynamique de leurs rendements montrent l'insuffisance de certains modèles adoptés, jusqu'à une période récente, par les sociétés d'assurance. L'examen des données historiques relatives aux rendements des actions permet de constater, par exemple, que l'hypothèse d'une distribution normale ne permet pas de prévoir des valeurs extrêmes de rendement tel que réalisé dans le passé (cf. Ahlgrim et al. [2005] et Mandelbrot [2005]). A ce niveau, certains chercheurs proposent l'adaptation d'un modèle de changement de régime dans lequel les rendements des actions peuvent être simulés sous l'un des deux régimes suivants : un premier régime avec une hypothèse de volatilité relativement faible et une moyenne relativement élevée des rendements des actions et un deuxième régime avec une hypothèse de volatilité relativement élevée et une moyenne relativement faible de ces rendements. Une probabilité de transition entre ces deux régimes peut être déterminée *a priori* (cf. Hardy [2001]). D'autres alternatives sont proposées notamment dans le cadre des modèles discontinus (cf. Merton [1976]).

Concernant les autres classes d'actifs, tels que l'immobilier et les produits dérivés, malgré les différents cadres d'hypothèses proposés pour leur projection, elles se heurtent souvent aux problèmes d'un historique peu profond, d'une liquidité insuffisante et de données confidentielles (cas des fonds de couverture). En matière de projection de grandeurs réelles sur le long terme, ces actifs sont souvent traités avec prudence et ne suscitent pas la grande part de l'intérêt des décideurs, en particulier dans le cas de l'allocation stratégique d'actifs d'un fonds de retraite où la priorité est souvent donnée aux actions, aux obligations et au monétaire (cf. Campbell et al. [2001]). Ceci ne remet pas en cause le potentiel que présentent ces actifs en tant que source de performance et/ou de couverture supplémentaire pour le portefeuille financier de la société (cf. Ahlgrim et al. [2005]).

### 2.2. Littérature sur les GSE

La littérature sur les GSE est abondante. On propose ici de classer les différents modèles en fonction de la structure de dépendance entre les variables, en distinguant deux catégories : structure par cascade et structure basée sur les corrélations. De même, pour chacun des modèles cités, on précisera l'objectif qui lui était associé lors de sa conception. A ce titre, on note que l'utilisation d'un GSE a souvent pour finalité soit la projection sur le long terme et la prise de décision dans le cadre de la gestion des risques (dans ce cas, l'intérêt porte sur des grandeurs et des valeurs réelles) soit l'évaluation des prix d'équilibre des produits financiers sur le court terme, dit *pricing*, afin de déterminer la stratégie de marché convenable (achat, vente, etc.).

#### 2.1.1. Modèles à structure par cascade

Une structure par cascade est définie comme une structure dans laquelle on en part de la détermination de la valeur d'une variable (par exemple l'inflation) pour ensuite déduire les valeurs des autres variables (taux réels, rendements des actions, etc.). Jusqu'au début des années 1980, les travaux académiques traitent souvent une partie seulement du problème rencontré par les actuaires : les travaux sont concentrés sur chacune des classes d'actifs financiers (les actions, les taux d'intérêt, l'inflation,..) indépendamment des éventuelles interactions entre elles. L'intégration de ces interactions et l'étude du choix du modèle de dynamique des actifs financiers devient nécessaire pour garantir la cohérence des projections par rapport à un contexte donné.

Le travail de Wilkie [1986] marque de ce point de vue un changement majeur. Il présente pour la première fois un modèle général qui inclut toutes les variables macro économiques et financières. Ce modèle a l'avantage d'être simple à implémenter, raison pour laquelle il est rapidement devenu populaire et considéré comme la référence de tous les modèles proposés durant les deux décennies postérieures, malgré ses nombreuses limites.

La première version du modèle de Wilkie a été appliquée dans le cadre de la mesure de la solvabilité d'une société d'assurance par la *Faculty of Actuaries* (1986). De même, un des premiers domaines d'application du modèle de Wilkie en actuariat était l'évaluation des engagements indexés sur les actions : dans ce cas on suppose que les prestations dépendent des prix futurs des actions et que la réserve est principalement investie en obligations. De façon générale, ce modèle est plutôt cohérent avec des logiques de besoin de capital et de projection de valeur (cas de la gestion actif-passif par exemple) qu'avec des logiques de *pricing*.

Wilkie se base sur une structure par cascade, telle que décrite ci-dessus : il postule que l'inflation est la variable indépendante – « la force motrice » – du modèle dont la détermination se fait en premier lieu pour ensuite en dériver les valeurs des autres variables, principalement les dividendes, les revenus de dividende, les taux d'intérêt et la croissance des salaires. Le graphique 1 illustre ce principe d'une structure par cascade. Dans ce cadre, Wilkie utilise un modèle autorégressif de premier ordre pour l'inflation. En 1995, il met à jour ce premier modèle en gardant les principes de sa structure par cascade mais en optant cette fois-ci pour une modélisation de l'inflation par un processus ARCH (*Autoregressive Conditional Heteroscedasticity*). Ceci a été justifié, selon Wilkie, par la capacité de ce type de processus à tenir compte des caractéristiques des distributions historiques des données observées sur le marché de la Royaume-Uni depuis 1919.

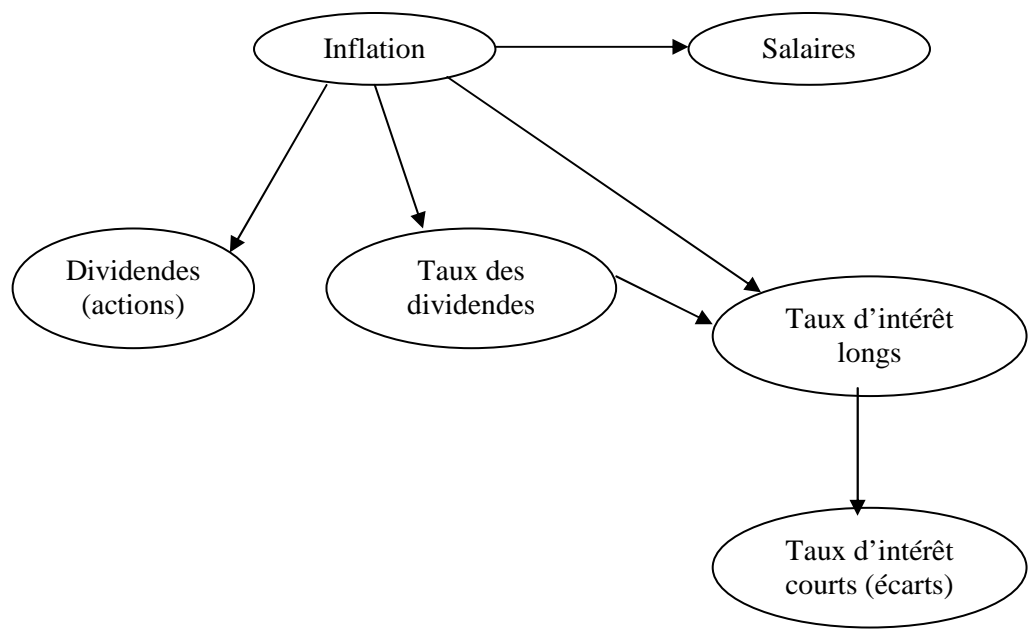

**Graph. Structure par cascade dans le modèle de Wilkie [1986]**

Toutefois, l'approche de Wilkie a depuis été remise en cause, notamment du fait de sa faible capacité prédictive : il s'agit d'un modèle utilisant un grand nombre de paramètres, dont l'estimation est délicate et qui empêche de fournir des projections pertinentes. Au surplus, le modèle de Wilkie se prête mal à l'évaluation des prix des actifs dérivés, ce qui constitue un handicap important. On pourra se référer sur ces points à Rambaruth [2003].

Les problèmes liés au modèle de Wilkie ont également été discuté par Daykin et Hey [1990] et Huber [1995] : certains des paramètres sont instables dans le temps et on constate une corrélation croisée significative entre les résidus des variables projetées. Le modèle des indices de prix pour l'inflation n'a pas de résidus normaux et ne permet pas la projection de périodes à chocs irréguliers avec des valeurs élevées de l'inflation. De même, la probabilité d'avoir des valeurs négatives de l'inflation avec ce modèle est élevée.

Par ailleurs, Ziemba et Mulvey [1998] décrivent un modèle de génération de scénarios économiques appelé CAP:Link (développé commercialement par la société Towers Perrin). Ce modèle est basé sur une structure par cascade de ces variables dont la force motrice est supposée être le taux d'intérêt nominal. Il est appliqué principalement dans la gestion actif-passif de long terme. Parmi les variables clés modélisées, on trouve l'inflation des prix et des salaires, les taux d'intérêts de différentes maturités (réels et nominaux), le taux de rendement et les taux de dividendes des actions et les taux de change.

Les variables financières sont déterminées simultanément pour différentes économies dans un cadre d'hypothèses générales. Ce modèle s'applique ainsi aux portefeuilles de pension et d'assurance. Les dynamiques des variables sont identiques pour tous les pays alors que les paramètres sont adaptés aux spécificités de chacun d'entre eux. Les auteurs soulignent que, de façon générale, les GSE remplissent au moins l'une de ces trois fonctions suivantes : la prévision, l'évaluation (*pricing*) et l'analyse du risque. Ils considèrent que l'élément clé d'un GSE est le modèle de taux d'intérêt et supposent donc que les taux longs et les taux courts sont corrélés à travers leurs termes de bruit blanc et que l'écart entre eux est contrôlé par un terme de stabilisation.

D'autres modèles de structure par cascade ont été développés pour l'Australie (cf. Carter [1991]), l'Afrique du Sud (cf. Thomson [1994]), le Japon (cf. Tanaka et al. [1995]) et la Finlande (cf. Ranne [1998]). L'élément commun de ces modèles réside donc dans le fait que le concepteur part de la spécification d'une structure en cascade du modèle à travers les hypothèses sur les liens de causalité entre les variables. En effet, cette structure en cascade permet un seul sens de causalité et exige du modélisateur le choix des liens les plus pertinents de point de vue économique. Par exemple, dans le modèle de Wilkie, la valeur de l'indice des prix permet de déduire la valeur de l'indice des salaires et non l'inverse. Le deuxième sens de causalité est supposé être faible (ou secondaire) sur le long terme.

**2.1.2. Modèles basés sur les corrélations**

La structure basée sur les corrélations repose quant à elle sur l'idée de permettre aux données disponibles (historiques) de déterminer une structure de corrélation simultanée entre les variables pour ensuite les modéliser et les calibrer en fonction de cette structure. Autrement dit, cette dernière est déterminée essentiellement à travers l'estimation des relations de dépendance observées simultanément dans le passé entre les variables modélisées (par exemple la corrélation linéaire observée dans le passé entre les rendements des actions et l'inflation est à retenir et à respecter lors de la projection dans le futur de ces variables). Ainsi, dans ce type de modèle, les données historiques disponibles sur les variables permettent de déduire la structure de dépendance entre elles. Les principaux modèles de GSE en littérature se sont basés sur cette structure.

En adoptant cette structure par corrélation, Campbell et al. [2001] présente une approche dont l'application a été effectuée dans le cadre de la détermination de l'allocation stratégique d'actifs pour un investisseur de long terme et en particulier les fonds de pension.

De son côté, Kouwenberg [2001] se base sur cette structure pour développer un modèle de génération de scénarios qui s'appuie sur un schéma d'arborescence pour la projection des scénarios. L'auteur compare l'effet du choix du schéma de projection sur l'allocation d'actifs optimale dans le cadre d'une gestion actif-passif d'un fonds de pension allemand. La structure d'arborescence retenue par Kouwenberg [2001] est plus adaptée à une série de modèles dynamiques de gestion actif-passif basées sur les techniques de programmation stochastique. Les caractéristiques de ce schéma de projection de scénarios sont présentées de façon détaillée à la section 3 de cet article.

Hibbert et al. [2001] présente un autre modèle qui génère des valeurs cohérentes, selon les auteurs, de la structure par terme des taux (taux nominaux, réels et d'inflation), des rendements des actions et des revenus de dividendes. Le modèle peut être utilisé pour générer des trajectoires potentielles de chacune de ces variables dans un cadre de modélisation financière et en considérant les différentes corrélations. Hibbert et al. [2001] fournit notamment une revue intéressante des taux d'intérêt, des taux d'inflation et des rendements des actions sur les cents dernières années. Leur modèle est présenté comme un outil de planification et de prise de décisions pour les investisseurs sur le long terme et non comme un outil d'évaluation des produits dérivés (ou *de pricing*).

Ahlgrim et al. [2005] propose enfin un modèle de GSE qui a le mérite d'être soutenu par la la *Casualty Actuarial Society* (CAS) et de la *Society Of Actuaries* (SOA), deux associations professionnelles reconnues aux Etats-Unis. Ahlgrim et al. [2005] partent essentiellement de la critique de deux points du modèle de Wilkie [1995] : la relation entre l'inflation et les taux d'intérêt est jugée incohérente et le traitement des rendements des actions par une approche autorégressive semble trop simplificateur au regard de l'historique observé. Ils proposent des processus alternatifs en justifiant leurs choix par des *backtesting*[9] sur des données historiques profondes. Le modèle d'Ahlgrim et al. [2005] rejoint le modèle de Hibbert et al. [2001] en se présentant comme un modèle de projection de valeurs sur le long terme et de gestion des risques.

D'autres modèles sont basés sur l'hypothèse que les marchés peuvent être soit faiblement efficients (c'est-à-dire les prix sur le marché reflètent toutes les informations relatives aux prix antérieurs de l'actif) soit fortement efficients (c'est-à-dire les prix sur le marché reflètent toutes les informations disponibles sur l'actif). De telles approches sont souvent appropriées pour la modélisation de court terme, en particulier pour des fins d'évaluation de produits dérivés. Pour le long terme, l'approche a moins de valeur, puisqu'elle ne tient pas compte des fondamentaux macro-économiques. Smith [1996] et Dyson et Exley [1995] présentent des modèles basés sur les principes du marché efficient pour le cas de la Grande Bretagne. En particulier, ce type de modèle cherche souvent à exclure les opportunités d'arbitrage et souvent ne suppose pas un retour à la moyenne pour les rendements de ces variables.

Une critique commune à tous les modèles ci-dessus, à structure par corrélation, est qu'ils sont fortement dépendants des données sur lesquelles ils sont basés. Autrement, si les rendements futurs relatifs à chaque variable du modèle possèdent des caractéristiques différentes de celles observées sur la période historique d'estimation, le GSE pourra conduire à des projections non pertinentes. La prise en compte des avis subjectifs des experts sur le marché (sociétés de gestion, banques d'investissement, etc.) pour fixer ces niveaux futurs de dépendance constitue une source alternative d'alimentation de ces modèles.

---

[9] Le *Backtesting* est le test d'une stratégie sur le passé et sur un panel d'actifs financiers.

On le voit, le panorama des modèles proposés dans la littérature est large ; on peut toutefois synthétiser quelques-uns de ces travaux comme suit:

| Structure  Objectif | Cascade | Corrélation |
|---|---|---|
| Projection et Gestion des risques | Wilkie [1986, 1995]  Ziemba et Mulvey [1998] | Campbell et al. [2001]  Kouwenberg [2001]  Hibbert et al. [2001]  Ahlgrim et al. [2005] |
| Evaluation (*pricing*) | - | Smith [1996]  Dyson et Exley [1995] |

Après avoir choisi la structure théorique du GSE, on arrive à l'étape de sa mise en œuvre pratique. Cette étape nécessite elle aussi des choix à effectuer que ce soit au niveau de la calibration des différents paramètres du GSE ou au niveau de la génération des trajectoires possibles de ses variables. Concernant la calibration d'un GSE, différentes techniques peuvent être citées se basant sur les données historiques, les données de marché ou les avis des experts. Le détail de ces techniques dépasse le cadre de cet article (pour plus de détails, voir Hibbert et al. [2001]). La section suivante s'intéresse à certaines problématiques liées à la génération des scénarios.

## 3. Mise en œuvre de la génération de scénarios dans un GSE

Deux questions principales peuvent être posées lors de la mise en œuvre d'un GSE : d'une part, sous quelle forme schématique doit-on représenter l'évolution dans le temps des scénarios futurs des variables financières et macro-économiques (inflation, rendement des actions,…) ? Et d'autre part, quelle méthodologie doit-on adopter pour la génération de ces scénarios ? La première question concerne la structure de projection des scénarios futurs des différentes variables, tandis que la deuxième a trait au choix du modèle d'évolution des valeurs des variables du GSE (processus stochastique, *Bootstrapping*, etc.) : on note que ces deux éléments sont cependant liés. Cette section vise à faire l'inventaire (non exhaustif) des différentes possibilités offertes face à ces deux problématiques.

En effet, comme mentionné au début de l'article, notre objectif sera de mettre en évidence les principaux éléments qui participent à l'amélioration de la qualité d'un GSE. Les états de sortie de ce dernier influencent directement les décisions prises en matière de gestion des risques ou d'évaluation des produits financiers. La détermination de la structure de projection des scénarios et le choix de la méthodologie de leur génération présentent deux étapes inévitables lors de la construction d'un GSE. Ils interviennent, en particulier, au niveau de la mise en œuvre opérationnelle du GSE, d'où l'intérêt de les étudier de façon détaillée et séparée.

### 3.1. Structure schématique de projection de scénarios pour un GSE

Comme mentionné ci-dessus, le choix de la structure de projection, appelée aussi structure schématique de projection, pour chacune des variables du GSE peut être considéré comme une problématique à part. Elle concerne la définition du schéma graphique de transition entre deux valeurs successives, observées à la date *t* et *t+1*, de la même variable. Ainsi à chaque variable du GSE peut correspondre une structure de projection particulière. Afin de simplifier l'analyse, on suppose dans la suite que la structure de projection choisie est la même pour toutes les variables. De même, on définit un nœud comme la réalisation possible de la variable modélisée à une date donnée. Une trajectoire correspond ainsi à l'ensemble des nœuds successifs qui forment un scénario futur possible d'évolution de la variable financière ou macro-économique.

Dans cette section, on présente, avec détails, les caractéristiques des structures de projection les plus utilisées en pratique. En particulier, deux principales structures peuvent être avancées à ce stade : la structure de projection linéaire (cf. Ahlgrim [2005]) d'une part et la structure de projection d'arbre (ou d'arborescence) d'autre part (cf. Kouwenberg [2001]). La différence principale entre ces deux structures de projection se situe au niveau de la nature de la dépendance entre les différentes trajectoires simulées. Alors que pour les structures de projection linéaire, une seule trajectoire est dérivée à partir de chaque nœud, les structures par arborescence supposent quant à elles que chaque nœud possède différents nœud-enfants et ainsi différentes trajectoires possibles sont déduites à partir de chaque nœud.

Par exemple, on se positionne dans le cas où la projection des rendements des actions se déroule sur deux périodes seulement et que pour les deux structures on obtient $n_1$ nœuds (ou rendements) à la fin de la première période. Si on opte pour une structure linéaire de projection, on ne peut obtenir que $n_1$ nœuds (ou rendements) à la fin de la deuxième période, chacun d'entre eux forme avec le nœud précédent une trajectoire distincte. Si par contre on opte pour une structure d'arbre, le nombre de scénarios à la fin de la deuxième période est $m_1$, avec $m_1$ supérieur à $n_1$ comme on projette différents rendements à partir de chacun des $n_1$ nœuds simulés fin de la première période.

Le graphique suivant illustre la différence entre ces deux structures :

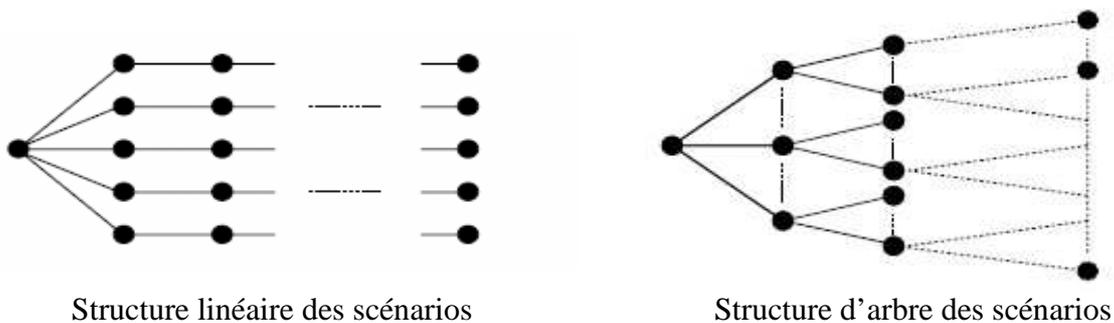

Structure linéaire des scénarios          Structure d'arbre des scénarios

Source Kouwenberg [2001]

La structure d'arbre des scénarios est la forme la plus récente et la plus complexe à utiliser. On focalisera notre intérêt dans la suite sur ce type de structure : il s'agit d'une structure plus adaptée que la structure linéaire pour l'application des techniques d'optimisation dynamique, en particulier dans le cas de la détermination de l'allocation stratégique d'actifs optimale (cf. Kouwenberg [2001]). En effet, chaque niveau dans l'arbre représente une date future (ou un moment de prise de décision) et les différents nœuds à chaque niveau représentent les réalisations possibles de la variable modélisée à cette date.

Le graphique suivant représente un exemple plus détaillé de cette structure :

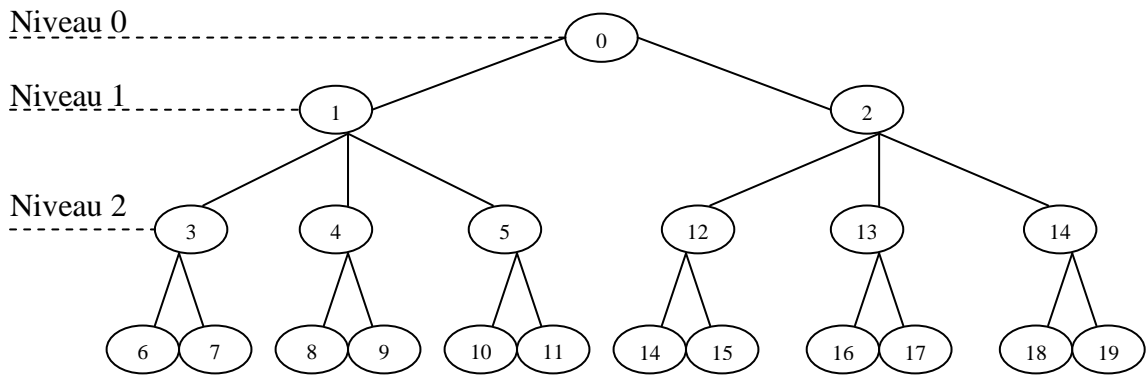

Comme il est montré dans ce graphique, le nombre des nœuds-enfants à chaque niveau n'est pas nécessairement égal à celui du niveau suivant. Par exemple, le nœud 0 dans le graphique a deux nœuds-enfants alors que les nœuds 1 et 2 ont trois nœuds-enfants. Deux niveaux de l'arbre peuvent ne pas présenter la même période de temps. Par exemple, dans le graphique ci-dessus, le niveau 0 peut représenter le début de l'année 0, le niveau 1 la fin de la deuxième année et le niveau 2 la fin de la dixième année. De même, dans certains arbres de scénarios complexes, tel que présenté ci-dessous, il pourrait avoir différents nombres de nœuds-enfants pour les nœuds d'un même niveau.

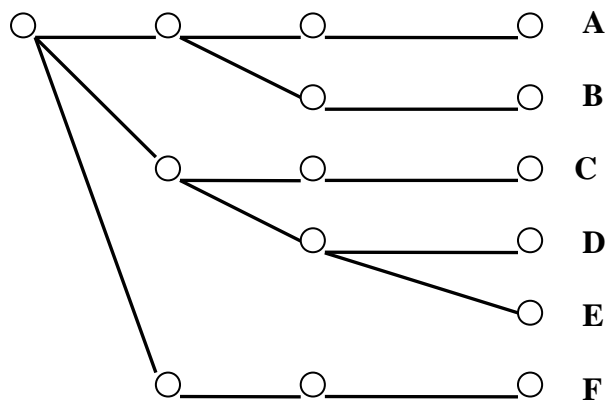

Il existe différentes représentations mathématiques possibles de l'arbre des scénarios. On se réfère ici à la formulation de Hochreiter et al. [2002].

Considérons d'abord un processus stochastique $(\xi_t)_{[t=0,1,...T]}$ discret dans le temps et continu dans l'espace et supposé représenter la dynamique des rendements des actions. L'analyse suivante peut être appliquée aux autres variables générées par le GSE. $\xi_0 = x_0$ représente la valeur d'aujourd'hui et elle est supposée être constante. La distribution de ce processus peut être le résultat d'une estimation, paramétrique ou non, basée sur des données historiques.

L'un des objectifs lors de la mise en place d'un GSE est de trouver un processus stochastique $\vec{\xi}_t$, qui prend seulement des valeurs finies et qui est le plus proche que possible du processus réel des rendements des actions : on parle dans ce cas de problème d'approximation. A titre d'hypothèse, le GSE est supposé avoir une structure de projection sous forme d'arbre pour ses différentes variables. On définit, pour cela, l'espace d'état fini de $\vec{\xi}_t$ par $S_t$ :

$$P\left\{\vec{\xi}_t \in S_t\right\} = 1$$

Soit $card\{S_t\}$ le cardinal de $S_t$. Si $x \in S_t$, on appelle le facteur de branchement de $x$, le nombre des noeuds issus directement de $x$, c'est à dire la quantité :

$$b(x,t) = card\left\{y : P\left\{\vec{\xi}_{t+1} = y, \vec{\xi}_t = x\right\} > 0\right\}$$

Intuitivement, le processus $(\vec{\xi}_t)_{[t=0,...,T]}$ peut être représenté sous forme d'arbre, avec comme racine le noeud $(x_0, 0)$. Les nœuds $(x,t)$ et $(y,t+1)$ sont connectés par un arc si $P\left\{\vec{\xi}_t = x, \vec{\xi}_{t+1} = y\right\} > 0$.

La collection de tous les facteurs de branchement $b(x,t)$ détermine la taille de l'arbre. Typiquement, on choisit le facteur de branchement avant et indépendamment de $x$. Dans ce cas, la structure de l'arbre est déterminée par le vecteur $[b(1), b(2), b(3),..., b(T)]$. Par exemple, un arbre [5, 3, 3, 2] a pour hauteur 4 et $1 + 5 + 5.3 + 5.3.3 + 5.3.3.2 = 156$ nœuds. Le nombre des arcs est toujours égal au nombre des nœuds moins 1. Dans ce cadre d'analyse, on peut considérer que la structure linéaire constitue un cas particulier de la structure d'arbre avec un facteur de branchement égal à 1 à partir de la deuxième composante du vecteur de la structure ci-dessus (c'est-à-dire [5, 1, 1, 1]).

Selon Dupacova et al. [2000], le problème d'approximation principal est un problème d'optimisation de l'un des deux types suivants et il est souvent fonction de la méthode de génération de scénarios retenue :

- Le problème à structure donnée (*The given-structure problem*) : quel processus discret $(\vec{\xi}_t)_{[t=0,...,T]}$ avec une structure de branchement $[b(1), b(2), b(3),..., b(T)]$ est le plus proche d'un processus donné $(\xi_t)_{[t=0,...,T]}$ ? Bien évidemment, la notion de proximité est à définir de manière appropriée.

- Le problème à structure libre (*The free-structure problem*) : ici aussi le processus $(\xi_t)_{[t=0,...,T]}$ est à approximer par $(\vec{\xi}_t)_{[t=0,...,T]}$ mais sa structure de branchement est à définir librement excepté que le nombre total des nœuds est fixé à l'avance. Ce problème d'optimisation hybride et combinatoire est plus complexe que le problème à structure donnée.

On note que la difficulté majeure lors de l'utilisation des arbres de scénarios est l'augmentation exponentielle dans le nombre de scénarios. Si trois scénarios sont générés pour chaque nœud à n'importe quel niveau parmi 21 niveaux par exemple, le nombre de scénarios générés sera $3^{21}$ (presque 3,5 milliards de scénarios).

Au delà du choix de la structure de projection pour les variables d'un GSE, l'analyse des approches possibles pour la détermination des valeurs futures des variables projetées, appelée aussi méthodologies de génération des scénarios, constitue un problème souvent rencontré par le constructeur d'un GSE. Le choix d'une méthodologie particulière n'est pas sans impact sur les résultats obtenus *in fine* (cf. Ahlgrim et al. [2008]).

### 3.2. Méthodologies pour la génération de scénarios économiques

Si on se place dans le même cadre d'analyse que celui de la sous-section précédente, on peut dire que cette partie vise à répondre à la question suivante: comment peut-on déterminer la valeur d'un nœud. Pour cela, différentes méthodologies, ayant pour finalité la génération de scénarios économiques, peuvent être trouvées dans la littérature (cf. Kaut et Wallace [2003] et Mitra [2006]). On propose dans ce travail de les classer en quatre groupes: les approches basées sur l'échantillonnage, les approches basées sur le *matching* des propriétés statistiques, les approches basées sur les techniques de *Bootstrapping* et les approches basées sur l'Analyse des Composantes Principales. Ce dernier groupe n'est pas indépendant des autres comme on va le voir dans ce qui suit.

#### 3.2.1. Les approches basées sur l'échantillonnage

Ces approches peuvent être classées en deux sous catégories : l'échantillonnage pur (ou traditionnel) et l'échantillonnage à partir de marginales et de corrélations spécifiées. Cette dernière a le mérite de générer des scénarios dans lesquels la corrélation entre les variables converge vers celle ciblée par le modélisateur.

L'échantillonnage pur est la méthode de génération de scénarios la plus connue. A chaque nœud de l'arbre de scénarios, on tire de façon aléatoire différentes valeurs à partir du processus stochastique $\{\xi_t\}$. Cela se fait soit par un tirage direct à partir de la distribution de $\{\xi_t\}$, soit par l'évolution du processus selon une formule explicite: $\xi_{t+1} = z(\xi_t, \varepsilon_t)$.

Dans ce cadre, la dynamique de plusieurs variables financières peut être supposée suivre un processus stochastique de type mouvement brownien géométrique et ses variantes. Les scénarios d'évolution de ces variables sont ainsi simulés à partir des hypothèses sur la discrétisation d'un processus brownien géométrique définit par exemple par:

$$dS(t) = \mu S(t)dt + \sigma S(t)dB(t)$$

$S(t)$ est le prix de l'actif, $\mu$ et $\sigma$ sont respectivement le *drift* et la volatilité. Le terme $dB(t)$ est un mouvement brownien, c'est-à-dire $B(t_2) - B(t_1) \sim N(0, |t_2 - t_1|)$. On pourra ainsi simuler des processus stochastiques sur un intervalle de temps donné, en attribuant des valeurs aléatoires au mouvement brownien et en calculant par la suite $S(t)$.

Les méthodes d'échantillonnage traditionnel permettent de constituer des échantillons seulement à partir d'une variable aléatoire univariée ; lorsqu'on veut tirer un vecteur aléatoire (correspondant à différentes variables), on aura besoin de tirer chaque composante marginale (chaque variable) de façon séparée pour les rassembler ensuite. Le résultant obtenu sera un vecteur de variables aléatoires indépendantes.

Concernant la convergence vers les moments statistiques souhaités (moyenne, variance, etc.), il existe différentes méthodes pour améliorer l'algorithme de l'échantillonnage pur. On pourra par exemple utiliser les méthodes de quadrature pour l'intégration ou les suites à discrépance faible (cf. Pennanen et al. [2002]). Pour les distributions symétriques on pourra utiliser les échantillonnages antithétiques. Une autre méthode pour améliorer la méthode d'échantillonnage pur est d'ajuster l'arbre obtenu de façon à avoir les valeurs cibles de la moyenne et de la variance (cf. Cariňo et al. [1994]).

Comme mentionné au début de cette partie, les méthodes d'échantillonnage traditionnel ont des limites au niveau de la génération des vecteurs multivariés, en particulier ceux avec une corrélation spécifiée. Cependant, il existe des méthodes qui résolvent ce problème en se basant sur des approches d'échantillonnage traditionnel pour ensuite ajuster la technique de contrôle de la corrélation entre les scénarios projetés. Ces méthodes, qu'on a appelé «échantillonnage à partir de marginales et de corrélations spécifiées» constituent donc une extension des approches d'échantillonnage traditionnel. La différence principale se situe dans le fait qu'elles permettent à l'utilisateur de spécifier à l'avance les distributions marginales ainsi que la matrice de corrélation cible. En général, il n'y a aucune restriction sur les distributions marginales, elles peuvent même appartenir à différentes familles.

A titre d'exemple, Deler et al. [2001] propose une méthode permettant de générer des variables aléatoires à partir des séries temporelles multi-variées $\{X_t; t=1,2,...\}$, où $X_t = (X_{1,t}, X_{2,t},..., X_{k,t})'$ est un vecteur aléatoire de dimension $(k \times 1)$ correspondant à l'observation à la date $t$ de ces variables. Pour cela, il construit un processus appelé processus de base $Z_t$ (assimilé à un vecteur auto régressif gaussien standard) et le transforme, à travers un système de translation de Johnson [1949], en un processus ayant au moins les quatre premiers moments (moyenne, écart-type, le coefficient de dissymétrie ou *skewness* et coefficient d'aplatissement ou *kurtosis*) du processus $X_t$. Il ajuste enfin la structure de corrélation de ce processus de base $Z_t$ de façon à obtenir celle du processus $X_t$. L'approche utilisée dans ce cas est celle de Deler et al. [2001a]. Elle se base sur la résolution d'un ensemble d'équations permettant de déterminer les corrélations à retenir entre les variables du vecteur $Z_t$ afin de refléter la structure de corrélation cible observée entre les variables du vecteur $X_t$. D'autres exemples de ces méthodes se trouvent dans Dupacova et al. [2009].

On note finalement que ces approches basées sur l'échantillonnage sont utilisées dans le cas où on a des hypothèses sur les fonctions de distribution des composantes marginales (ou des différentes variables modélisées).

### 3.2.2. Les approches basées sur le *matching* des propriétés statistiques

Dans les situations où il n'y a pas d'hypothèses sur la distribution marginale du processus de génération de scénarios, les approches basées sur le *matching* des propriétés statistiques, en particulier les moments, sont les plus adaptées. Un processus de génération des scénarios par le *matching* des moments s'intéresse souvent aux trois ou aux quatre premiers moments de chacune des projetées (moyenne, variance, skewness, kurtosis) ainsi qu'à la matrice de corrélation. Ces méthodes peuvent être étendues à d'autre propriétés statistiques (tel que les quantiles, etc.). Le générateur de scénarios par le *matching* des moments va ensuite construire une distribution discrète satisfaisant les propriétés statistiques sélectionnées. Par exemple, Hoyland et al. [2003] commence par spécifier le nombre minimal de scénarios qu'il faut générer pour ensuite obtenir l'arbre de scénarios par une optimisation non linéaire, où l'objectif est de minimiser l'erreur entre les moments théoriques de la variable et ceux fournis par l'arbre.

### 3.2.3. Les approches basées sur les techniques de *Bootstrapping*

Il s'agit des approches les plus simples pour générer des scénarios en utilisant seulement les données historiques disponibles sans aucune modélisation *a priori* de la dynamique d'évolution des variables du GSE (cf. Albeanu et al. [2007]). Le *Bootstrapping* se base essentiellement sur la constitution d'échantillons à partir des données observées. Dans ce cadre, les valeurs de chaque scénario représentent un échantillon de rendements d'actifs obtenu par un tirage aléatoire de certains rendements observés déjà dans le passé. Par exemple, afin de générer des scénarios de rendement sur les dix prochaines années, un échantillon de 120 rendements mensuels tirés aléatoirement sur les 120 rendements des dix dernières années est utilisé. Ce processus est répété un certain nombre de fois afin de générer plusieurs scénarios possibles dans le futur. Autrement, dans le cas où on a $k$ actifs à projeter, et en supposant que l'on dispose d'un historique de $p$ périodes, on tire avec remise un entier entre $1$ et $p$ et on prend toutes les valeurs des $k$ actifs à cette même date de façon à tenir compte de la corrélation historique entre ces derniers.

### 3.2.4. L'utilisation de l'Analyse en Composantes Principales

L'Analyse des Composantes Principales (ACP) est une méthode générique d'analyse de données ayant plusieurs dimensions (cf. Bouroche et al. [1980]). Elle permet l'identification des facteurs clés régissant les tendances de ces données et de réduire leur dimension tout en conservant le plus d'information possible. Pour cela, l'ACP se base sur l'identification des vecteurs propres, des valeurs propres et des covariances. En effet, il s'agit d'une méthode descriptive qui dépend d'un modèle géométrique plutôt que d'un modèle probabiliste. L'ACP propose de réduire la dimension d'un ensemble des données (échantillon) en trouvant un nouvel ensemble de variables plus petit que l'ensemble original des variables, qui néanmoins contient la plupart de l'information de l'échantillon. Autrement, pour un ensemble de données dans un espace à $N$ dimensions, on recherche un sous-espace à $k$ dimensions (défini par $k$ variables) tel que la projection des données dans ce sous-espace minimise la perte d'information. Ces $k$ variables seront appelés composantes principales et les axes qu'elles déterminent axes principaux. L'implémentation numérique de la méthode ACP est ainsi accessible. En pratique, l'ACP a pour objet de réduire le nombre des variables de départ du modèle pour ensuite appliquer une approche traditionnelle de génération de scénarios (parmi celles proposées ci-dessus) pour les composantes principales.

## 4. Mesure de la qualité d'un GSE

La mesure de la qualité d'un GSE peut avoir deux caractères : qualitatif et quantitatif. Dans les deux cas, l'objectif sera de garantir des états de sortie du modèle qui permettent de prendre les meilleures décisions.

### 4.1. Mesure qualitative de la qualité d'un GSE

Pour Hibbert et al. [2001], les propriétés qu'un « bon » GSE doit avoir sont les suivantes : la représentativité, la plausibilité économique, la parcimonie, la transparence et l'évolution.

Le modèle doit « imiter » le comportement des actifs financiers dans le monde réel en captant leurs principales caractéristiques (représentativité). De même, les différents scénarios générés devront être plausibles et raisonnables pour les experts du marché financier. Cela passe par l'étude de la forme de la distribution et des corrélations des différentes variables du modèle.

A ce titre, le degré de correspondance des résultats aux données historiques constitue un champ d'investigation intéressant, malgré les critiques contre l'hypothèse de reproduction des événements historiques dans le futur. L'utilité de la comparaison des résultats obtenus par rapport aux observations déjà réalisées sur le marché peut varier en fonction des attentes du gestionnaire en termes d'évolution future du contexte macro-économique. En fait, l'objectif de similitude entre les deux séries (projetées et historiques) peut être abandonné si on juge par exemple un changement profond du contexte économique par rapport au passé (exemple : modification importante des poids des économies à l'échelle internationale, explosion de l'endettement public,…).

Le comportement joint des différentes variables simulées dans les scénarios doit aussi présenter un niveau suffisant de plausibilité et de cohérence en respectant les principes économiques. Cependant, il est à noter qu'il n'y a pas de consensus sur certaines propriétés importantes des actifs financiers. L'exemple du principe de retour à la moyenne sur le marché des actions, qui stipule que les rendements sur le long terme des actions convergent vers une tendance donnée, en est une illustration (cf. Hibbert et al. [2001]).

La parcimonie fait référence, quant à elle, à la simplicité des modèles proposés par le GSE, leur permettant d'être d'une part compréhensibles par les utilisateurs et d'autre part implémentables sur des supports informatiques. Par ailleurs, la parcimonie permet une meilleure capacité prospective.

La transparence est enfin un élément déterminant pour le succès du modèle, fortement lié à son potentiel de communication : les résultats obtenus devront être présentés sous forme de graphiques clairs et détaillés. Finalement, Le potentiel du modèle à évoluer et à élargir le champ de son étude est un élément d'attractivité supplémentaire.

Pour Zenios [2005], les critères qu'un GSE doit satisfaire afin de garantir sa bonne qualité sont au nombre de trois : l'exactitude (*correctness*), la précision (*accuracy*) et la cohérence (*consistency*).

Les scénarios devront avoir des propriétés qui sont **justifiées et soutenues par la recherche académique**. Par exemple, la structure par terme devra refléter le phénomène de retour à la moyenne des taux d'intérêt constaté dans plusieurs études universitaires et techniques (cf. Vasicek [1977] et Hibbert et al.[2001]). De même, la structure par terme pourra être déduite à partir des changements dans le niveau, la pente et la courbure tels que mentionné dans certaines études économiques (cf. Heath et al. [1992]).

De même, afin de vérifier la condition d'exactitude, les scénarios devront couvrir les scénarios importants observés dans le passé ainsi que tenir compte des **évènements qui ne sont pas observés avant,** mais qui ont de fortes chances d'être observés sous les conditions actuelles de marché.

Comme dans plusieurs cas, les scénarios représentent une discrétisation d'un scénario continu, l'accumulation d'un nombre d'erreurs dans le discrétisation est inévitable. Différentes approches peuvent être utilisées pour assurer que l'échantillon de scénarios représente la fonction de distribution continue sous jacente.

La précision est assurée lorsque **le premier et le plus grand moment des scénarios convergent vers ceux de la distribution théorique sous-jacente**. (le *matching* des moments et des propriétés statistiques sont souvent utilisés afin d'assurer que les scénarios gardent les moments théoriques de la distribution qu'ils représentent).

La demande de précision peut mener à la génération d'un nombre élevé de scénarios. Ceci dans le but de créer une discrétisation fine de la distribution continue et afin d'accomplir la précision qu'on considère appropriée et acceptable pour le problème en question. Cependant, on note qu'il ne faut perdre de vue que les erreurs de modèle et de calibrage conduisent à ne pas consacrer trop d'énergie à l'obtention d'un niveau de précision, formellement satisfaisant mais sans intérêt de point de vue opérationnel.

Lorsque les scénarios sont générés pour différents instruments (par exemple, les obligations, la structure par terme, etc.), il est important de voir que les scénarios sont cohérents en interne. Par exemple, les scénarios dans lesquels une augmentation dans le taux d'intérêt tient lieu simultanément avec une hausse des prix des obligations sont théoriquement non cohérents (cf. Ahlgrim et al. [2008]), même si à une échelle individuelle chacun des deux scénarios, celui du taux et celui des prix, peut être avoir lieu étant donné certains facteurs exogènes (par exemple, une augmentation des taux par la banque centrale accompagnée par l'augmentation de la demande sur les obligations comme actifs refuges sur le marché). La prise en compte de la **corrélation** entre les différents instruments peut être utilisé afin d'assurer la cohérence des scénarios.

Voici des exemples d'études de la qualité de certaines méthodologies de génération de scénarios présentées ci-dessus:

Concernant les approches basées sur l'échantillonnage : elles permettent de probabiliser un ensemble de scénarios futurs en tenant compte de la possibilité de la reproduction des scénarios passés. Si le choix de la dynamique des variables est justifié par des références académiques, ces approches assurent la condition d'exactitude. Par ailleurs, les conditions de précision et de cohérence peuvent aussi être préservées : avec un choix de techniques de calibration et de discrétisation adéquates (précision) et en tenant compte de la structure de corrélation entre les variables dans les modèles (cohérence).

On peut noter que le *matching des moments* assure par définition la condition de précision puisqu'elle garde les propriétés des moments. De même, le *matching* des matrices de covariance assure la cohérence des scénarios. Cependant, cette approche est assez générale et n'est pas validée par des études académiques ce qui est nécessaire pour que l'approche respecte la condition d'exactitude.

De son coté le bootstapping des données historiques préserve la corrélation observée, mais ne satisfait pas la condition d'exactitude pour la génération de scénarios car il ne suggèrera pas un rendement qui n'est jamais observé dans le passé (ce qui est contre-intuitif, comme la reproduction du passé ne constitue qu'une partie des scénarios possibles dans le futur et non la totalité des scénarios potentiels, confirmation donnée par la crise économique actuelle qui a donné lieu à des rendements records jamais observés au paravent). Lorsqu'ils sont tirés de façon appropriée, les scénarios de cette approche satisfont les conditions de précision et de cohérence puisqu'ils sont en parfaite cohérence avec les observations réelles du passé.

### 4.2. Mesure quantitative de la qualité d'un GSE

Dans cette partie, inspirée des travaux de Kaut et Wallace [2003], on essaye de présenter certains critères permettant de tester de façon quantitative la qualité d'un GSE non pas en tant qu'outil de génération de la distribution réelle de ses variables mais plutôt en tant qu'outil d'aide à la prise de décision. L'intérêt dans ce travail sera porté sur la performance des scénarios comme un *input* au modèle de prise de décision et sur la mesure de la qualité des décisions subséquentes que proposent ces scénarios (stabilité, absence de biais, etc.). L'objectif n'est donc pas la mesure du degré de correspondance des scénarios projetés par rapport à la distribution réelle des variables du GSE, une distribution qui reste inconnue quelque soit les outils d'estimation et d'approximation employés. On note aussi que la connaissance de la distribution réelle supprime le besoin de passer par un GSE pour prévoir les valeurs futures des variables financières et macro-économiques.

#### 4.2.1. Présentation du problème

Les notations utilisées dans cette partie sont identiques à celles utilisées précédemment.[10]. On s'intéresse à la fonction objectif d'un modèle d'optimisation synthétisé comme suit :

$$\min_{x \in X} F(x, \xi_t)$$

où $\xi_t$ correspond donc à un processus stochastique discret dans le temps et continu dans l'espace.

La résolution directe du problème réel (théorique) étant supposée être impossible, le passage par un processus discret $\{\vec{\xi}_t\}$ dont l'évolution est représentée sous forme d'un arbre ne constitue qu'une solution d'approximation (empirique). Une fois effectué, la fonction objectif devient :

$$\min_{x \in X} F(x, \vec{\xi}_t)$$

Comme mentionné ci-dessus, on se propose dans ce travail de lier la performance du GSE étudié à la qualité des décisions qu'il permet d'obtenir et non pas à la performance statistique d'approximation et d'estimation du processus réel. D'autre part, on rappel que l'erreur d'estimation des paramètres ne fait pas partie de l'objet de cet article ce qui ne réduit en aucun cas l'importance d'étudier la robustesse des décisions obtenues par le GSE par rapport aux choix effectués pour ces paramètres (cf. Meucci [2005]).

---

[10] Le choix de la structure de projection n'influence pas l'analyse menée dans cette sous section.

L'erreur, au niveau décisionnel, de l'approximation d'un processus stochastique $\{\xi_t\}$ par une discrétisation $\{\vec{\xi}_t\}$, peut être définit, soit comme la différence entre les valeurs optimales de la fonction objectif des deux problèmes : le problème réel (théorique) et le problème d'approximation (empirique), soit comme la différence entre les solutions proposées par ces deux mêmes problèmes.

Le premier type d'erreur, lié à la valeur optimale de la fonction objectif, peut être formulé comme suit :

$$e_f(\xi_t, \vec{\xi}_t) = F(\arg\min_x F(x; \vec{\xi}_t); \xi_t) - F(\arg\min_x F(x; \xi_t); \xi_t)$$

$$= F(\arg\min_x F(x; \vec{\xi}_t); \xi_t) - \min_x F(x; \xi_t) \qquad (1)$$

On note que $e_f(\xi_t, \vec{\xi}_t) \geq 0$, étant donné que le deuxième élément de l'écart est le vrai minimum, alors que le premier est la valeur de la fonction objectif à une solution approximée. La comparaison est effectuée non pas au niveau des solutions optimales $(x^*)$ mais plutôt au niveau des valeurs de la fonction objectif correspondantes. En fait, notre intérêt porte en premier lieu sur la valeur minimale et non pas sur les différentes solutions qui permettent *in-fine* d'avoir cette valeur (appelée aussi la performance).

Il y a deux problèmes rencontrés dans l'équation ci-dessus de $e_f(\xi_t, \vec{\xi}_t)$ :

- trouver la valeur « réelle» de $F(x; \vec{\xi}_t)$ pour une solution donnée $x$.
- trouver la valeur « réelle » de $F(x; \xi_t)$ qui suppose la détermination de la solution réelle $x^*$.

Alors que le second problème est souvent difficile à résoudre, car il nécessite la résolution du problème d'optimisation avec le processus continu réel, le premier problème peut être résolu, à travers la simulation en temps discret par exemple. Dans la sous section suivante, on présente différents indicateurs de mesure de l'erreur lié à la qualité des décisions obtenues par un GSE donné. Ces indicateurs se basent sur la définition de conditions à satisfaire par la solution empirique.

#### 4.2.2. Le test de la qualité des décisions obtenues par le GSE

Il existe au moins deux conditions nécessaires à satisfaire par un GSE afin de garantir la fiabilité des décisions déduites à partir de ses projections. La première condition est **la stabilité** de la valeur de la fonction objectif optimale par rapport aux différents arbres de scénarios: c'est-à-dire si on génère plusieurs arbres (avec les mêmes paramètres initiaux : tendance, volatilité, etc.) et on résout le problème d'optimisation pour chacun d'eux, on devrait avoir les mêmes valeurs optimales de la fonction objectif (mêmes performances). L'autre condition est que l'arbre de scénarios n'introduit **aucun biais** au niveau de la vraie solution $(x^*)$ et ainsi la solution proposée empiriquement doit correspondre à la solution optimale réelle. La première condition peut dans certaine mesure être testée alors qu'un test direct de la deuxième est souvent considéré comme impossible étant donné que la fonction réelle est inconnue.

### 4.2.2.1. La condition de stabilité

On note que deux types de stabilité peuvent être avancés : la stabilité interne de l'échantillon et la stabilité externe de l'échantillon. Ces deux types d'indicateurs interviennent, dans cet article, au niveau de la valeur optimale de la fonction objectif et non au niveau de la solution optimale du problème.

En fait, la stabilité interne de l'échantillon signifie que si on génère $K$ arbres de scénarios avec la discrétisation $\{\vec{\xi}_t\}$ d'un processus donné $\{\xi_t\}$, et on résout le problème d'optimisation pour chacun de ces arbre on devrait avoir (approximativement) la même valeur optimale de la fonction objectif avec comme solutions optimales $x_k^*$, $k=1...K$. Autrement : $F(x_k^*;\vec{\xi}_{tk}) \approx F(x_l^*;\vec{\xi}_{tl}) \qquad k,l \in 1...K$

La stabilité externe de l'échantillon concerne quant à elle les performances des arbres de la distribution réelle et elle est définit comme suit :

$$F(x_k^*;\xi_t) \approx F(x_l^*;\xi_t) \qquad k,l \in 1...K$$

Une autre formulation possible de ces deux indicateurs sera:

La stabilité interne : $\min_{x} F(x;\vec{\xi}_{tk}) \approx \min_{x} F(x;\vec{\xi}_{tl})$

La stabilité externe : $F\left(\underset{x}{argmin}\, F(x;\vec{\xi}_{tk});\xi_t\right) \approx F\left(\underset{x}{argmin}\, F(x;\vec{\xi}_{tl});\xi_t\right)$

En utilisant l'équation (1) de la fonction d'erreur, on déduit que pour le cas de la stabilité externe: $e_f(\xi_t,\vec{\xi}_{tk}) = e_f(\xi_t,\vec{\xi}_{tl})$

Il y a une différence importante entre les deux définitions : alors que pour la stabilité interne de l'échantillon on a besoin seulement de résoudre le problème d'optimisation basé sur les scénarios projeté par le GSE, la mesure de la stabilité externe passe par l'évaluation de la fonction objectif réelle $F(x;\xi_t)$. En pratique, on ne peut pas tester précisément la stabilité externe comme on n'a pas la distribution réelle des variables projetées (rendements, taux, etc.).

Cela peut être démontré par l'exemple uni-périodique et uni-dimensionnel suivant[11] illustrant la différence entre les deux types d'indicateurs :

$$\min_{x \in R} F(x;\xi) = E^{\xi}\left[(x-\xi)^2\right]$$

---

[11] Pour simplifier, $\xi$ désigne $\xi_t$ dans cet exemple.

Ce problème peut être résolu théoriquement et avoir sa solution explicite pour n'importe quelle distribution de $\xi$ :

$$\begin{aligned}
F(x;\xi) &= E\left[(\xi-x)^2\right] \\
&= E\left[\left((\xi-E[\xi])+(E[\xi]-x)\right)^2\right] \\
&= E\left[(\xi-E[\xi])^2\right] + E\left[2(\xi-E[\xi])(E[\xi]-x)\right] + E\left[(E[\xi]-x)^2\right] \\
&= Var[\xi] + 0 + (x-E[\xi])^2
\end{aligned}$$

et la solution optimale est alors :

$$x^* = \underset{x\in\Re}{argmin}\, F(x;\xi) = E[\xi]$$

$$F(x^*;\xi) = \underset{x\in\Re}{min}\, F(x;\xi) = Var[\xi]$$

Cette solution est dite explicite parce que son obtention ne nécessite pas la simulation de la distribution de $\xi$ et elle constitue du fait une solution générale et déterministe. Considérons le cas dans lequel on génère des arbres de scénarios $\vec{\xi}_k$ ($k=1...K$) et qu'on obtient les solutions $x_k^* = E\left[\vec{\xi}_k\right]$.

On suppose dans un premier temps que la méthode de génération de scénarios est telle que tous les échantillons $\vec{\xi}_k$ ont d'une part la même moyenne réelle $E[\vec{\xi}_k]=E[\xi]$ et d'autre part des variances $Var[\vec{\xi}_k]$ différentes. Ainsi $F(x_k^*;\vec{\xi}_k) = Var[\vec{\xi}_k]$ sera différente d'un échantillon à l'autre d'où une non stabilité interne. En même temps, $x_k^* = x^*$, donc $F(x_k^*;\xi) = F(x^*;\xi) = Var[\xi]$, ce qui confirme une stabilité externe de l'échantillon.

Si par contre, on suppose que la méthode de génération de scénarios produit des échantillons ayant tous la même variance réelle ($Var[\vec{\xi}_k]=Var[\xi]$), mais des moyennes différentes, on obtient dans ce cas $F(x_k^*;\vec{\xi}_k) = Var[\vec{\xi}_k] = Var[\xi]$ et la stabilité interne de l'échantillon est ainsi vérifiée. De l'autre côté, la valeur de $F(x_k^*;\xi) = Var[\xi] + (E[\vec{\xi}_k]-E[\xi])^2$ diffère entre les échantillons et la résolution du problème ne vérifie pas la stabilité externe de l'échantillon.

La question qui peut être posée à ce stade est laquelle des deux types de stabilité est la plus importante pour tester la méthode de génération de scénarios. Avoir la stabilité externe sans avoir la stabilité interne signifie que la performance obtenue par approximation peut ne pas correspondre à la performance réelle. Le cas opposé, c'est-à-dire stabilité interne sans la stabilité externe, est quant à lui plus gênant pour le processus décisionnel, puisque la performance réelle des solutions empiriques dépendra de l'arbre de scénarios réel sélectionné, et on ne pourra pas ainsi trancher au niveau de la meilleure solution obtenue par approximation alors que dans le premier cas on avait au moins une solution empirique stable théoriquement mais dont la performance empirique reste non stable, ce qui paraît *a priori* moins gênant.

Une autre question possible est l'étude d'une stabilité interne de l'échantillon non pas au niveau de la performance mais au niveau des solutions elles même : ainsi on cherche à avoir la même solution optimale quelque soit l'arbre de scénarios généré. Dans l'exemple ci-dessus, on peut remarquer qu'il est possible d'avoir une non stabilité interne de la valeur optimale de la fonction objectif (différentes variances empiriques $Var[\vec{\xi}_k]$) tout en gardant une stabilité interne au niveau des solutions elles mêmes (les solutions $x_k^* = E[\xi]$ sont les mêmes pour tous les arbres). Ceci permet de déduire immédiatement la stabilité externe de l'échantillon. Ainsi, si on détecte une stabilité interne de l'échantillon au niveau de la fonction objectif, on devrait regarder aussi au niveau des solutions. Par contre l'autre sens d'analyse n'est pas toujours vrai et on peut avoir une stabilité externe même si la stabilité interne au niveau des solutions n'est pas vérifiée.

On peut conclure *in-fine* que la stabilité est une condition minimale à respecter par la méthode de génération de scénarios. Ainsi, avant de commencer à travailler avec un nouveau modèle d'optimisation, ou une nouvelle méthode de génération de scénarios, il sera judicieux d'effectuer les tests de stabilité : le test de la stabilité interne de l'échantillon et dans la mesure du possible le test de la stabilité externe de l'échantillon.

### 4.2.3.2. La condition d'absence de biais

En plus de la condition de stabilité (à la fois interne et externe de l'échantillon), la méthode de génération de scénarios ne devra pas introduire de biais au niveau de la solution proposée. Autrement, la solution du problème d'approximation basé sur les scénarios fournis par notre GSE:

$$\vec{x}* = \underset{x}{argmin}\, F(x;\vec{\xi}_t)$$

devra être une solution optimale du problème original. Ainsi, la valeur de la vraie fonction objectif pour la solution obtenue, $F(\vec{x}*;\xi_t)$, devra être approximativement égale à la valeur optimale réelle $\underset{x}{min}\, F(x;\xi_t)$ :

$$F(\vec{x}*;\xi_t) = F\left(arg\, \underset{x}{min}\, F(x;\vec{\xi}_t);\xi_t\right) \approx \underset{x}{min}\, F(x;\xi_t)$$

En utilisant l'équation (1) de la fonction d'erreur, cette condition s'exprime come suit : $e_f(\xi_t,\vec{\xi}_t) \approx 0$.

Le problème est que le test de cette propriété n'est pas possible dans la plupart des cas, étant donné que cela nécessite la résolution du problème d'optimisation avec le processus continu réel, supposé jusque là comme inconnu.

## 5. Application numérique

Afin d'illustrer les points développés tout au long de cet article, on présente dans ce qui suit un exemple numérique prenant le cas d'un investisseur (compagnie d'assurance par exemple) intéressé par la répartition de son capital entre $n=3$ actifs financiers sur le marché européen : les actions, les obligations et le monétaire. Chacun de ces trois classes d'actifs sera représenté par un indice dont la série de données historiques mensuelles, entre le 01/01/1999 et le 31/10/2009, est récupérée à partir de la base de données Bloomberg : ainsi on a retenu l'indice EONIA pour le monétaire, l'indice obligataire EFFAS-Euro pour les obligations et l'indice DJ Eurostoxx50 pour les actions. On suppose aussi que ce problème d'optimisation de portefeuille est mono-périodique : la projection des valeurs de rendement de chacun des trois actifs se fait sur une seule période (par exemple une année).

Au niveau du choix de **la méthodologie de génération des scénarios** de notre GSE, on associe un processus stochastique reflétant une hypothèse de normalité des rendements à chacun des trois indices, soit :

$$\mu_i + \sigma_i \varepsilon_{t,i} \qquad \text{avec } i=1,2,3$$

La tendance (ou la moyenne annualisée) $\mu_i$ et la volatilité $\sigma_i$ des rendements des trois indices sont estimées sur une base historique (elles correspondent respectivement à la moyenne annualisée et à la volatilité empiriques de chacune des trois séries de données retenues). $\varepsilon_{t,i}$ suit une loi normale de moyenne nulle et de variance égale à 1.

**La structure de dépendance entre les variables de notre GSE** est bien une structure par corrélation. Les trois indices sont simultanément dépendants l'un de l'autre lors de leur projection et le niveau des corrélations est déterminé à travers les corrélations fixées en amont entre leurs bruits $\varepsilon_{t,i}$ ($i=1,2,3$).

**La fonction objectif** de notre investisseur sera définit comme suit :

$$Min \quad F(R_p; \varepsilon_t) = \sigma(R_p)$$
$$tel\ que\ \ p \in U$$
$$E(R_p) \geq m_0$$

Avec $R_p$ le rendement total du portefeuille composé des trois actifs (monétaire, obligations et actions), $E(R_p)$ la valeur espérée du rendement du portefeuille, $\sigma(R_p)$ la volatilité du rendement du portefeuille et $m_0$ la valeur minimale du rendement espéré du portefeuille (égale à 4 % dans notre cas). En pratique, l'espérance sera estimée par la moyenne empirique. Rappelons que pour déduire le rendement du portefeuille sur un période donnée, on utilise la formule suivante :

$$R_p = \sum_{j=1}^{n=3} p_j R^j$$

Avec $p_j$ et $R^j$ correspondants respectivement au poids et au rendement de l'actif $j$ sur la même période.

$U$ correspond à l'univers des allocations (ou compositions de portefeuille) à tester. La détermination de cet univers passe par la fixation d'un pas, par exemple de 5 %, pour les poids des différents actifs. Autrement :

$$U = \left\{ (p_1,...,p_n) \setminus \sum_{j=1}^{n} p_j = 1, 1 \geq p_j = c \times 0,05 \geq 0, \forall j \in \{1,...,n\}, c \in \{0,1,...,20\} \right\}$$

Dans notre cas, 3 actifs et un pas de 5 %, on obtient 231 allocations à tester.

$F(R_p ; \varepsilon_t)$ correspond à la fonction objectif réelle présentée dans la section précédente comme $F(x ; \xi_t)$[12].

**La structure de projection** adoptée dans cette application est une structure linéaire et ce afin de simplifier la méthode de résolution de la fonction objectif. Cette structure peut être illustrée par exemple par le graphique suivant correspondant à la variable rendement des actions projetée sur une période de 20 ans avec 500 trajectoires :

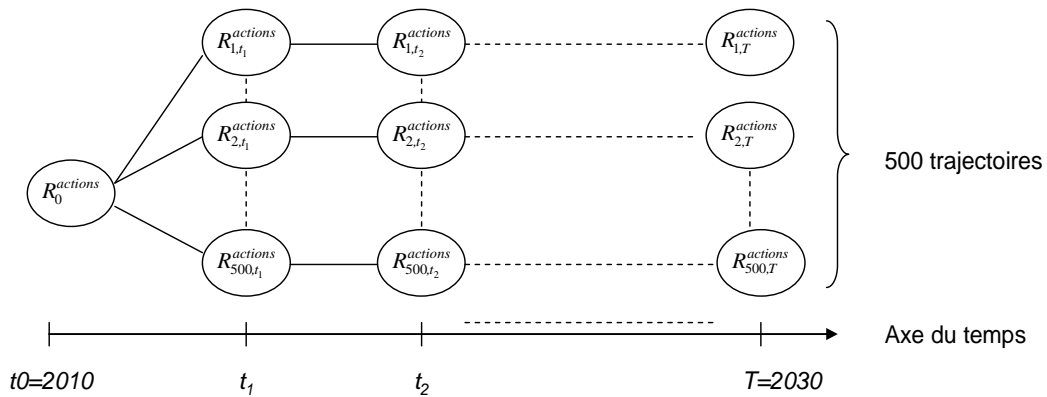

avec $R_{i,t}^{actions}$ le rendement de l'indice des actions entre la date $t$ et $t-1$ à la $i^{ème}$ trajectoire (ou scénario).

La taille de la structure de projection sera définie comme le nombre de trajectoires simulées dans cette structure. Un ensemble de scénarios désignera l'ensemble des trajectoires obtenues d'une variable dans le cas d'une structure de projection linéaire (équivalent de l'arbre de scénarios dans le cas de structure par arborescence).

**Dans notre étude on envisage de procéder comme suit :** On fixe différentes tailles des structures de projection adoptées dans notre GSE : par exemple 50, 1000, 5000 et 10 000 trajectoires. Pour chacune de ces valeurs, on fait tourner notre GSE $m$=30 fois en résolvant à chaque fois le problème d'optimisation présenté ci-dessus. On obtient ainsi 120 valeurs optimales empiriques de la fonction objectif (30 valeurs pour chacune des 4 tailles proposées). Les valeurs de ces différents paramètres (taille, nombre de fois qu'on fait tourner notre GSE) sont choisies de façon arbitraire. L'idée est de tester la stabilité et la distribution de la valeur optimale empirique de la fonction objectif ainsi que de la solution empirique optimale (c'est-à-dire l'allocation optimale d'actifs) par rapport à la taille de la structure. Cette étude essayera de juger *in fine* la qualité des solutions fournies par notre GSE pour le preneur de décision.

---

[12] La caractérisation « réelle » vient du fait que dans notre cas la résolution du problème passe par la simulation et la projection de scénarios stochastiques pour les différents actifs : on parle dans ce cas d'une résolution empirique basée sur la méthode Monte Carlo et non d'une résolution réelle déterminée généralement à l'aide de formules explicites.

Avant d'exposer et d'analyser les résultats obtenus, on rappelle que la stabilité interne de l'échantillon nécessite que, pour une taille de structure donnée, les valeurs optimales de la fonction objectif ne dépendent pas de l'ensemble de scénarios projetés. L'idéal sera donc, dans ce cas, d'avoir, pour une taille de structure donnée, 30 valeurs optimales approximativement identiques de la fonction objectif. Pour la stabilité externe de l'échantillon, les valeurs réelles de la fonction objectif (la variance) doivent être approximativement les mêmes pour les différents ensembles de scénarios projetées. De même, ces valeurs devront être approximativement égales à celles obtenues dans le test de stabilité interne. Comme mentionné précédemment, cette stabilité externe de l'échantillon peut être déduite à partir de la stabilité interne de l'échantillon au niveau de la solution empirique retrouvée (l'allocation optimale dans notre cas).

Dans le cadre du test de la stabilité interne de l'échantillon, le tableau suivant présente les statistiques relatives aux 30 valeurs optimales de la fonction objectif et ce pour les différentes tailles proposées.

**Tableau : Propriétés statistiques des valeurs de la fonction objectif pour les différentes tailles retenues des structures de projection de scénarios**

| Description du test | | | Nombre de scénarios | | | |
|---|---|---|---|---|---|---|
| Type de test | Fonction objectif | Valeur | **50** | **1000** | **5000** | **10 000** |
| Interne | $F(R_p;\bar{\varepsilon}_k)$ | Moyenne  Ecart-type | 0,0288  0,0050 | 0,0275  0,0011 | 0,0275  0,0007 | 0,0277  0,0005 |

De même, le graphique ci-dessous illustre la distribution de ces valeurs optimales (médiane, quantile 25% et 75% ainsi que le minimum et le maximum sur les 30 arbres de chaque taille).

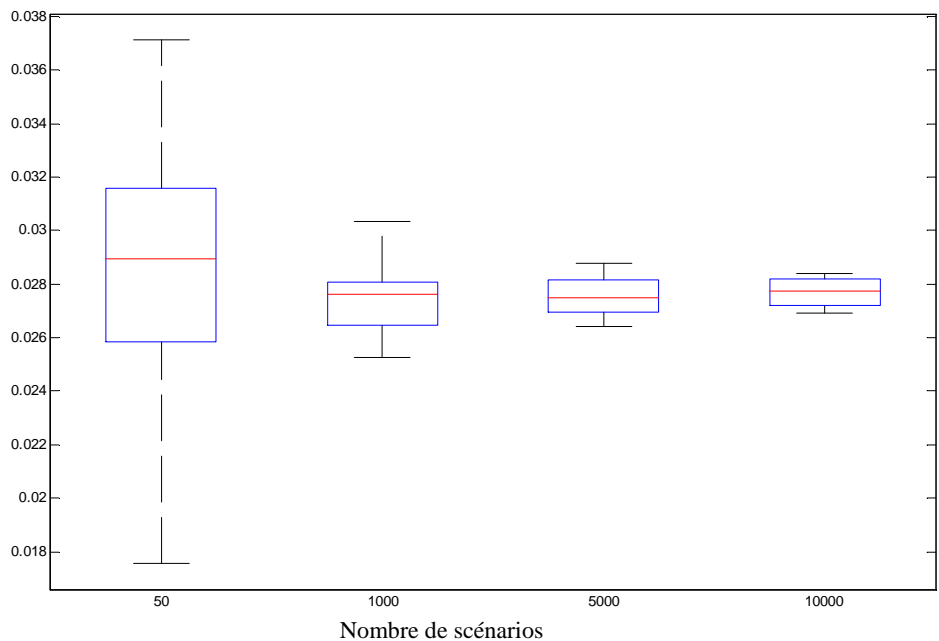

**Graph.** *Box plot* **de la distribution des valeurs de la fonction objectif pour différentes tailles de scénarios**

Comme on peut le constater, plus le nombre de scénarios augmente plus la valeur optimale de la fonction objectif a tendance à baisser et à se stabiliser autour d'une valeur centrale (en particulier à partir de 5 000 scénarios). La condition de stabilité interne de l'échantillon peut ainsi être respectée par notre GSE à partir d'une certaine taille de la structure.

D'un autre côté, si la variation du poids optimal de chaque actif est relativement faible (bornée entre deux valeurs proches) cela garantie la stabilité externe de l'échantillon. Dans ce cadre, on analyse les 30 vecteurs d'allocations optimales obtenus pour les ensembles de scénarios ayant pour taille 10 000. Les statistiques de ces solutions sont présentées dans le tableau suivant, ainsi que le graphique correspondant ci-dessous, et permettent de constater relativement une stabilité externe de l'échantillon (vu que les écarts types et/ou les étendus des poids des différents indices sont relativement faibles).

**Tableau : Propriétés statistiques des poids des 30 allocations optimales obtenues pour la taille de 10 000 scénarios**

|  | Moyenne | Ecart-type | Etendu | Minimum | Maximum |
|---|---|---|---|---|---|
| Monétaire | 0,37 | 0,0252 | 0,05 | 0,35 | 0,4 |
| Obligations | 0,55 | 0,0497 | 0,1 | 0,5 | 0,6 |
| Actions | 0,08 | 0,0254 | 0,05 | 0,05 | 0,1 |

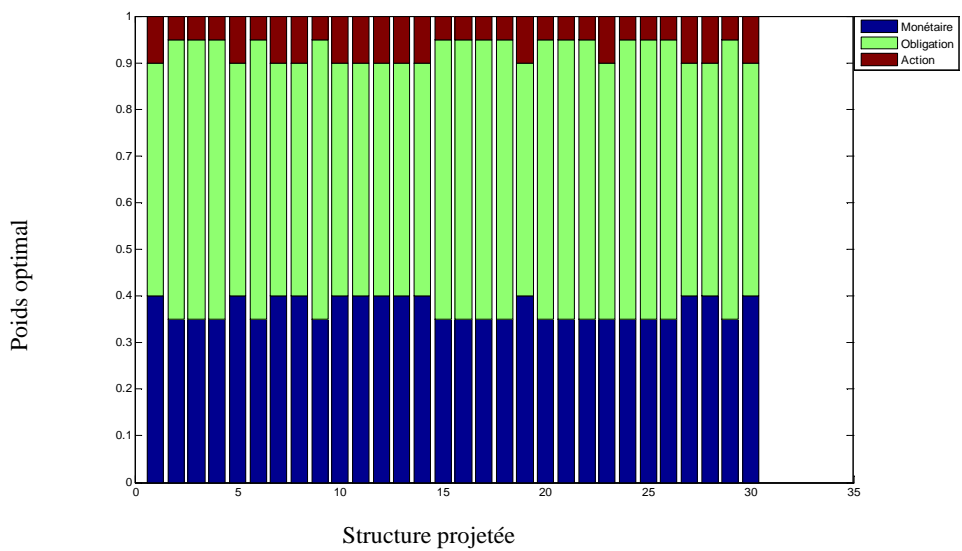

**Graph. Les allocations optimales obtenues pour les différents ensembles de scénarios (30 ensembles) ayant chacun la taille de 10 000 scénarios**

Les résultats obtenus à travers cette application constituent des indicateurs des quels le décideur peut s'en servir pour justifier la stabilité de sa décision finale par rapport à son générateur de scénarios : les graphiques et les tableaux présentés confirment la possibilité d'avoir une stabilité interne et une stabilité externe de l'échantillon. En pratique, les hypothèses de travail et les variables de risque prises en compte dans le processus décisionnel sont évidemment beaucoup plus nombreuses et plus compliquées mais cela ne limite pas l'intérêt de ces indicateurs comme des outils d'aide à la mesure de la performance de la décision.

# 6. Conclusion[13]

Le glissement sémantique du « modèle d'actifs » vers le « générateur de scénarios économiques » matérialise la profonde évolution de ce sujet depuis une dizaine d'années maintenant.

Longtemps utilisée dans le cadre d'études spécifiques et ponctuelles, la construction d'une représentation des actifs dans lequel l'assureur peut investir ses avoirs et du contexte économique dans lequel ils évoluent est devenue un élément essentiel de la description des risques que celui-ci doit gérer.

Que ce soit dans le cadre prudentiel, pour la détermination des provisions et du capital de solvabilité, pour sa communication vers les tiers (*embedded value* et états comptables) ou pour ses besoins de pilotage techniques (choix d'allocations stratégique et tactique, tests de rentabilité, etc.), l'organisme assureur doit disposer d'un cadre rigoureux et cohérent prenant en compte l'ensemble des actifs de son bilan et les risques associés.

Si l'identification de ces risques peut être considérée comme relativement aboutie, la crise financière a mis en évidence certaines faiblesses dans leur modélisation. Deux éléments sont ainsi mis en évidence et vont sans doute donner lieu à de nombreux développements dans les prochaines années :

---

[13] Voir Planchet et al. [2009].

- la structure de dépendance entre les actifs ;
- le risque de liquidité, intimement associé à la gestion efficace des couvertures financières.

La quasi-totalité des modèles actuels s'appuient sur des structures de dépendance dans lesquels la corrélation tient une place centrale ; de nombreux travaux, dont on pourra trouver une synthèse dans Kharoubi-Rakotomalala [2008], mettent en évidence le caractère dynamique de l'intensité de la dépendance. En pratique, l'intensité de la dépendance augmente dans les situations défavorables, ce qui limite l'efficacité des mesures de diversifications calibrées avec des structures ne prenant pas en compte cet effet. L'introduction de structures de dépendance non linéaires intégrant de la dépendance de queue apparaît ainsi comme un élément incontournable de l'évolution des générateurs de scénarios économiques.

Le risque de liquidité est également apparu comme un élément majeur de la crise des *subprimes* et, plus généralement, de la crise financière qu'elle a engendrée. Au moment où la généralisation des approches *market consistent* impose d'évaluer les options et garanties financières des portefeuilles dans la logique de détermination du coût de leur couverture, encore faut-il pour que le montant obtenu ait du sens que la couverture puisse être réajustée régulièrement, ce qui n'est possible qu'avec des actifs liquides. Cela impose par conséquent si ce n'est une adaptation de l'approche dans le cas d'actifs peu liquides, à tout le moins la prise en compte d'une prime de liquidité pour refléter dans le montant affiché ce risque d'impossibilité de gérer idéalement la couverture.

Au-delà de ces deux éléments structurants, l'efficacité opérationnelle des modèles mis en œuvre dépend, on l'a vu, dans une très large mesure de la pertinence des paramètres retenus pour les alimenter. La détermination de ses paramètres est complexe et fait appel à la fois à des considérations d'ordre statistique (exploitation des historiques), économique (cohérence des valeurs de long terme prédites par le modèle avec les relations économiques fondamentales), financières enfin (cohérence avec les prix observés sur le marché). La prise en compte rationnelle de ces différentes composantes nécessite une réflexion spécifique et fait partie intégrante des choix structurants en termes de gestion des risques que peut effectuer l'organisme assureur.

Enfin, la définition d'indicateurs de performance du modèle contribue une meilleure alimentation du processus de prise de décision. La mise en évidence de l'exploitation de ces indicateurs dans le cas de modèles plus compliqués et plus concrets, tel que les modèles de gestion actif-passif, pourrait constituer un champ d'investigation des travaux futurs.

On le voit, les sujets de réflexions ne manquent pas pour développer des modèles plus robustes et mieux à même de rendre compte de la complexité des risques économiques et financiers supportés par un organisme assureur.

# 7. Bibliographie